\documentclass[%
 reprint,
 amsmath,amssymb,
 aps,
showkeys,
]{revtex4-2}

\usepackage{graphicx}%
\usepackage{color}
\usepackage{dcolumn}
\usepackage{bm}
\usepackage[utf8]{inputenc}
\usepackage{multirow}
\usepackage{booktabs}
\usepackage{physics}
\usepackage{capekCommands}
\usepackage[nolist]{acronym}
\usepackage{subcaption}

\newcommand{\subto}{\T{s.t.}}

\newacro{QCQP}{quadratically constrained quadratic program}

\begin{document}

\title{Performance Bounds of Magnetic Traps for Neutral Particles}

\author{Jakub Liska}
\email{jakub.liska@fel.cvut.cz}
\author{Lukas Jelinek}
\author{Miloslav Capek}
\affiliation{Department of Electromagnetic Field, Czech Technical University in Prague, Prague, Czech Republic}

\date{\today}

\begin{abstract}
Knowledge of the fundamental limitations on a magnetic trap for neutral particles is of paramount interest to designers as it allows for the rapid assessment of the feasibility of specific trap requirements or the quality of a given design. In this paper, performance limitations are defined for convexity of magnetic trapping potential and bias field using a local approximation in the trapping center. As an example, the fundamental bounds are computed for current supporting regions in the form of a spherical shell, a cylindrical region, and a box. A Pareto-optimal set considering both objectives is found and compared with known designs of the Baseball trap and Ioffe-Pritchard trap. The comparison reveals a significant gap in the performance of classical trap designs from the fundamental limitations. This indicates a possibility of improved trap designs and modern techniques of shape synthesis are applied in order to prove their existence. The topologically optimized traps perform almost two times better as compared to conventional designs. Last, but not least, the developed framework might serve as a prototype for the formulation of fundamental limitations on plasma confinement in a wider sense.
\end{abstract}

\maketitle

\section{Introduction}
Magnetic trapping of neutral particles prepared in low-field-seeking Zeeman's states is a well-documented topic~\cite{Wing_NeutralPartTrapQuasistaElMagFields,Gov_MagTrapNeutralParticles,PerezRios_HowMagTrapWork} with major applications in preparing the grounds for subsequent optical cooling~\cite{Chu_ManipulationNeutralPart-Lecture,CohenTannoudji_ManipulAtomsWithFotons-Lecture,Phillips_LaserCoolingTrapNetralAtoms_Lecture, Phillips_LaserCoolElMagTrapNeutralAtoms} and Bose-Einstein condensation~\cite{Anderson_BoseEinsteinCondDiluteAtomicVapor,Davis_BoseEinsteinCondGasNaAtoms, Bradley_BoseEinsteinCondAtomicGasAttraciveInteract}. In the current state of the art, several successful designs of traps are used and thoroughly described~\cite{Bergeman_MagStatTrapFielsNeutralAtoms}. Among the most popular designs are the Ioffe-Pritchard trap~\cite{Bergeman_MagStatTrapFielsNeutralAtoms,Pritchard_CoolNeutralAtomsMagTrapPrecSpectroscopy,Gott_ResultsConfMagTraps} and a trap shaped like the seams of a baseball~\cite{Bergeman_MagStatTrapFielsNeutralAtoms,BaseballMagneticField1966,Hiskes_WhoMadeBaseball}, which are able to trap particles of temperature $T \approx 6.3 \cdot 10^{-3} I m/k_\T{B} \, \T{K}$  with $I$ being the current flowing in the conductors, $m$ being the magnitude of the magnetic dipole moment, and $k_\T{B}$ being Boltzmann's constant~\cite{Bergeman_MagStatTrapFielsNeutralAtoms}.

Magnetic traps are characterized by many performance metrics and criteria~\cite{Yang_DevelopmentHighFieldSupercondIoffeMagTraps,Ahokas_LargeOctupoleMagTrapResearchAtomicHydrogen, Wing_NeutralPartTrapQuasistaElMagFields,Gov_MagTrapNeutralParticles,PerezRios_HowMagTrapWork,Bergeman_MagStatTrapFielsNeutralAtoms,Gott_ResultsConfMagTraps,Yang_DevelopmentHighFieldSupercondIoffeMagTraps,Harris_MagTrapsNearlyUntrappableParticlesDevelopmentHighFieldSupercondIoffeTraps}. The most important are trapping depth, trapping volume, magnetic field magnitude in the trapping center, dissipated power, thermal management and physical forces. In this paper, the studied metrics are the depth of the trap and the field magnitude inside the trap which suppresses the probability of spin-flip leading to the loss of trapped particles~\cite{Gov_MagTrapNeutralParticles}. The performance is normalized with respect to the dissipated power, which is related to the thermal management and physical forces.

Although the designs of existing traps are advanced and sophisticated~\cite{Yang_DevelopmentHighFieldSupercondIoffeMagTraps,Ahokas_LargeOctupoleMagTrapResearchAtomicHydrogen,Harris_MagTrapsNearlyUntrappableParticlesDevelopmentHighFieldSupercondIoffeTraps}, it is important to ask how they perform in comparison to an ideal trap. This question is addressed in this paper given the building material of the current-carrying region and having complete freedom when shaping stationary current density in it. The optimal currents give fundamental bounds against which the realized designs should be compared when judging their performance, as well as show which current paths lead to high-quality traps.

The fundamental bounds proposed in this paper are based on an idea originally used to bound the performance of
antennas~\cite{GustafssonTayliEhrenborgEtAl_AntennaCurrentOptimizationUsingMatlabAndCVX,GustafssonCismasuJonsson_PhysicalBoundsAndOptimalCurrentsOnAntennas_TAP,JelinekCapek_OptimalCurrentsOnArbitrarilyShapedSurfaces} and recently broadened to cover extremal
optical cross-sections~\cite{2020_Gustafsson_NJP,2021_Jelinek_OPEX} or radiative heat transfer~\cite{Venkataram_FundamentalLimitsToRadiativeHeatTransferPRL,Molesky_FundamentalLimitsToRadiativeHeatTransferPRB}, see~\cite{2021_Chao_Arxiv} for a recent review of the topic and an extensive list of references. The basis of this framework is the use of field quantities as optimized variables and is typically connected to electromagnetic field descriptions via field integral equations~\cite{ChewTongHu_IntegralEquationMethodsForElectromagneticAndElasticWaves,Harrington_TimeHarmonicElmagField,Schwinger_ClassicalElectrodynamics}, and quadratically constrained quadratic program~\cite{NocedalWright_NumericalOptimization,BoydVandenberghe_ConvexOptimization}.

\section{Formulation}~\label{sec:form}
Neglecting the effect of gravity, which is to be treated later, trapping of a low-field seeking particle is, within the adiabatic approximation~\cite{Gov_1DtoyModelMagTrapping,Gov_MagTrapNeutralParticles}, described by magnetic potential energy~\cite{Zangwill_Modern_Electrodynamics,Bergeman_MagStatTrapFielsNeutralAtoms}
\begin{equation}
\label{eq:pot}
    \phi = - \V{m} \cdot \V{B} \approx m |\V{B}|.
\end{equation}
The approximation assumes a time-averaged magnetic dipole moment~$\V{m}$
\begin{equation}\label{eq:mPol}
    \V{m} \approx - m \frac{\V{B}}{|\V{B}|}
\end{equation}
of the particle released into the magnetic trap pointing antiparallel in the direction of the magnetic field.

As is common in the literature~\cite{PerezRios_HowMagTrapWork,Weinstein_MicroscopicMagneticTrapsNeutralAtoms,Gov_MagTrapNeutralParticles,Bergeman_MagStatTrapFielsNeutralAtoms}, the quadratic approximation is used to quantify the properties of the magnetic potential well forming the trap.
\begin{figure*}
    \centering
     \begin{subfigure}[b]{8.6cm}
         \centering
         \includegraphics{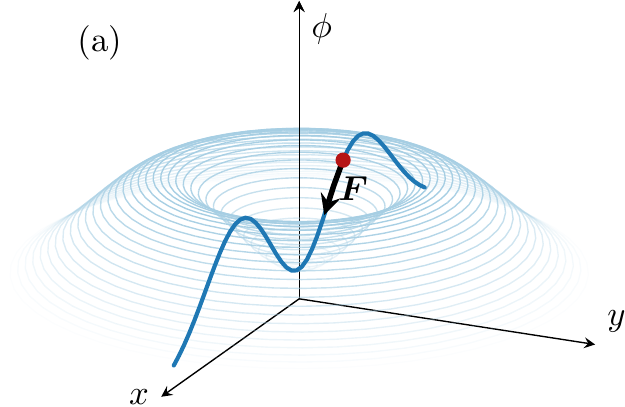}
     \end{subfigure}
     \hfill
      \begin{subfigure}[b]{5.1cm}
         \centering
         \includegraphics{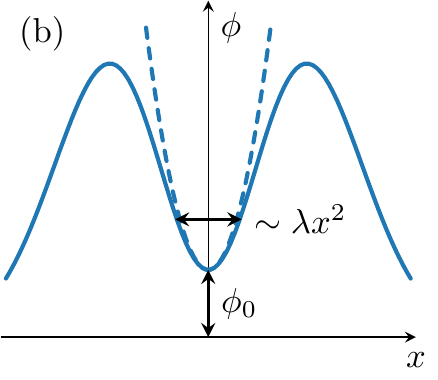}
     \end{subfigure}
     \hfill
     \begin{subfigure}[b]{3.5cm}
         \centering
         \begin{subfigure}[b]{\textwidth}
         \centering
         \includegraphics{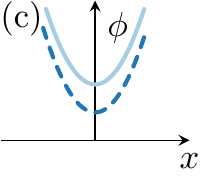}
     \end{subfigure}
     \vfill
      \begin{subfigure}[b]{3.5cm}
         \centering
         \includegraphics{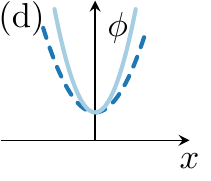}
     \end{subfigure}
     \end{subfigure}
    \caption{(a) Schematic of a two-dimensional isotropic potential well. (b) The cut along the $x$-axis from the left panel is approximated by a quadratic function at the trapping center where the force acting on the dipole is zero and $\phi_0$ is the potential energy in the trapping center. In this isotropic case, the curvature is identical in all directions and proportional to eigenvalue~$\lambda$. Panels (c, d) show a comparison of two potential wells with the same curvature (c) or the same potential energy (d), but different trapping performances. Solid lines show better performance.}
    \label{fig:schPotWells}
\end{figure*}
At that point, the potential~\eqref{eq:pot} is expanded around the center of the trap (chosen as the coordinate origin) as
\begin{equation}\label{eq:Taylor}
    \phi(\V{r}) \approx \phi_0 - \V{F} \cdot \V{r} + \frac{1}{2} \V{r} \cdot \M{H} \cdot \V{r},
\end{equation}
where $\V{F}$ is the magnetic force acting on the dipole~\cite{Zangwill_Modern_Electrodynamics}
\begin{equation}\label{eq:gradient}
    F_i = \V{m} \cdot \frac{\partial \V{B}}{\partial i} =- \frac{\partial \phi}{\partial i} \biggr\rvert_{\V{r}=\V{0}},
\end{equation}
and $\M{H}$ is a symmetric Hessian matrix (tensor) evaluated at the same point, the elements of which read
\begin{equation}\label{eq:hess}
    H_{ij} = - \dfrac{\partial F_i}{\partial j} = \frac{\partial^2 \phi}{\partial j \partial i} \biggr\rvert_{\V{r}=\V{0}},
\end{equation}
with~$\partial i \ \T{and} \ \partial j$ denoting derivatives with respect to Cartesian coordinates, \ie,~$i,j \in \left\{x,y,z \right\}$. The Hessian matrix characterizes the curvature of the potential at the trapping center. Specifically, its eigenvalues~$\lambda_i$
\begin{equation}
    \M{H} \V{q}_i = \lambda_i \V{q}_i, \quad i \in \{1,2,3\}
\end{equation}
determine the derivatives of the force (curvature of the potential) along the directions defined by the corresponding eigenvectors. Figure~\ref{fig:schPotWells} shows the properties of the quadratic approximation and demonstrates the meaning of the above given quantities.

In order to simplify the algebra and to prepare the grounds for establishing fundamental bounds on magnetic traps, auxiliary quantities quadratic in magnetic field are introduced as follows. The squared magnitude of the magnetic field is denoted as
\begin{equation}~\label{eq:potB}
    \Phi = \V{B} \cdot \V{B},
\end{equation}
its gradient is denoted~$\V{\Gamma}$ and defined component-wise as
\begin{equation}\label{eq:forceB}
    \Gamma_i = \frac{\partial \V{B}}{\partial i} \cdot \V{B} +  \V{B} \cdot \frac{\partial \V{B}}{\partial i},
\end{equation}
and the Hessian matrix of function~$\Phi$ is denoted as~$\M{\Xi}$ with elements
\begin{equation}
\label{eq:hessB}
    \Xi_{ij} = \frac{\partial^2 \V{B}}{\partial j \partial i} \cdot \V{B} +  \V{B} \cdot \frac{\partial^2 \V{B}}{\partial j \partial i} + \frac{\partial \V{B}}{\partial i} \cdot \frac{\partial \V{B}}{\partial j} + \frac{\partial \V{B}}{\partial j} \cdot \frac{\partial \V{B}}{\partial i}.
\end{equation}
All expressions are written in symmetric form to be prepared for the subsequent optimization process. With the help of~\eqref{eq:potB}--\eqref{eq:hessB}, and in the vicinity of trapping center~\eqref{eq:Taylor}, the interaction of the magnetic dipole with the external magnetic field can be written as
\begin{align}
        \label{eq:potTerm0}
        \phi &=  m \sqrt{\Phi}, \\
        F_i &= - \frac{m^2}{2 \phi} \Gamma_i, \\
        H_{ij} &=  \frac{m^2}{2 \phi} \Xi_{ij} - \frac{1}{\phi} F_i F_j. \label{eq:potTerm2}
\end{align}

The metrics defining quality of magnetic traps for neutral particles, see  Fig.~\ref{fig:schPotWells}, are assessed as follows:
\begin{itemize}
    \item The squared magnitude of magnetic field~$\Phi$ corresponds to the ability of the magnetic trap to avoid spin-flips and should be high enough.
    \item Magnetic force~$\V{F}$, acting upon a trapped particle, is required to be zero in all directions of the magnetic trap, which is the stationary condition for the trapped particle. The condition is satisfied if, and only if, gradient~$\V{\Gamma}$ is a zero vector.
    \item Eigenvalues~$\lambda_i$ of Hessian matrix~$\M{H}$ are desired to be positive so that the magnetic field forms a potential well in the trapping center. Since force~$\V{F}$ is a zero vector in the trapping center, only the first term in~\eqref{eq:potTerm2} is active and therefore the eigenvalues~$\xi_i$ of the Hessian matrix~$\M{\Xi}$ are required to be positive. For high localization of the trapped particle, the curvature of the potential should be high.
\end{itemize}

In the static approximation used, the evaluation of the above quantities is based on Biot-Savart's law~\cite[section~10.2]{Zangwill_Modern_Electrodynamics}, \begin{equation}
\label{eq:BiotSavart}
    \V{B}(\V{r}) = \frac{\MUE}{4 \pi} \int \limits_\varOmega  \dfrac{\V{J}(\V{r}') \times \left(\V{r} - \V{r}'\right)}{|\V{r} - \V{r}'|^3} \D V',
\end{equation}
where~$\V{J}$ is the current density flowing in the conductors of the trap ($\V{r} \in \varOmega$) and~$\MUE$ is the permeability of the vacuum. For a real design of a trap, current density~$\V{J}$ is unique and is given by the material distribution used and excitation, and~\eqref{eq:potTerm0}--\eqref{eq:potTerm2} can be directly used to judge the quality of the trap.

To set up the fundamental bounds on performance, the dependence on material distribution and excitation must be relaxed. To that point, consider current~$\V{J}$ as a free variable (impressed current in a vacuum). Optimizing over this variable, a question is then posed on what is the best possible current density existing in a given region which will lead to the best trapping properties. Quantities $\phi, \lambda_i$ are the objectives to be maximized (optimized) in the sense of fundamental bounds with a constraint $\Gamma_i = 0 \, \T{T}^2 \, \T{m}^{-1}, i \in \{x,y,z\}$. The worst case is always considered among the eigenvalues~$\lambda_i$,
\begin{equation}\label{eq:minEig}
    \lambda_\T{min} = \min \limits_i \lambda_i, \quad i \in \{1,2,3\}.
\end{equation}

An additional constraint preventing the current density to reach infinity in magnitude has to be set. A convenient one is the restriction on power lost~\cite{Zangwill_Modern_Electrodynamics} in the supporting current material
\begin{equation}
    P_\T{L} = \dfrac{1}{\sigma} \int \limits_\varOmega \left| \V{J} \left( \V{r} \right) \right|^2 \T{d}V,
\end{equation}
where~$\sigma$ is the conductivity of the material used. The lost power heats the conductors and it is constrained by the maximum allowed lost power~$P_\T{L}^\T{max}$.

Not including a normalization introduced later, the desired multi-objective optimization problem, setting the upper bounds on performance of any magnetic trap with current density~$\V{J}$, reads
\begin{equation}\label{eq:optim}
\begin{split}
    \max \limits_{\V{J}} \ & \xi_\T{min} \\
    \subto \ & P_\T{L} = P_\T{L}^\T{max}, \\
    & \Phi = \Phi_\T{s}, \ \Phi_\T{s} \in [\Phi_\T{min}, \Phi_\T{max}], \\
    & \Gamma_i = 0, \ i \in \{x,y,z\},
\end{split}
\end{equation}
where sweep in field strength~$\Phi_\T{s}$ provides the trade-off between field strength and potential convexity, thus forming a Pareto-optimal set~\cite{1978CohonMultiobjectiveProgrammingAndPlanning} of values~$\phi$ and $\lambda_\T{min}$.

The feasibility region is restricted by $\Phi_\T{max}$ above which the solution leads to concave potential in the trapping center or violates the maximum allowed lost power~$P_\T{L}^\T{max}$. The lower limit~$\Phi_\T{min}$ denotes the minimum field strength for which the adiabatic approximation~\cite{Gov_1DtoyModelMagTrapping,Gov_MagTrapNeutralParticles} assumed in~\eqref{eq:pot} holds. Fundamental bounds on the performance of traps with vanishing magnetic field at the trapping center, such as the quadrupole and hexapole traps, are therefore not studied here. These traps are nevertheless impractical with respect to particle spin flips.

\section{Results}
\label{sec:res}
The computation and evaluation of magnetic trap properties are based on current density. The technique used in this paper is Galerkin method~\cite{Kantorovich1982} or \ac{MoM}~\cite{Harrington_FieldComputationByMoM,Gibson_MoMinElectromagnetics}, see Appendix~\ref{ap:MoM}. This applies to realized traps, as well as to fundamental bounds. Conductors are assumed in the form of highly conducting surfaces~\cite{SenoirVolakis_ApproximativeBoundaryConditionsInEM}.

In the case of finding the optimal current distribution using~\eqref{eq:optim}, the chosen current supporting regions are a spherical shell, a cylindrical shell with a ratio of the diameter and height equal to one, further referenced only as a cylindrical region, and a box. While the choice is free (multilayered surface regions or volumetric current supporting regions can also be used), the considered canonical surfaces are advantageous in a sense of lower computational complexity~\footnote{Especially volumetric current supporting regions suffer from a large number of degrees of freedom and become extremely computationally demanding in the case of topology optimization.} and are based on geometries of practically used trap designs.

The fundamental bound corresponding to a particular current support then presents an upper limit on performance of any trap design that fits into this current supporting region. The above-mentioned regions then limit known trap designs, namely, the Ioffe-Pritchard trap is supported by a cylindrical region or by a box and the baseball trap is supported by a spherical shell.
\begin{figure}
    \centering
    \includegraphics{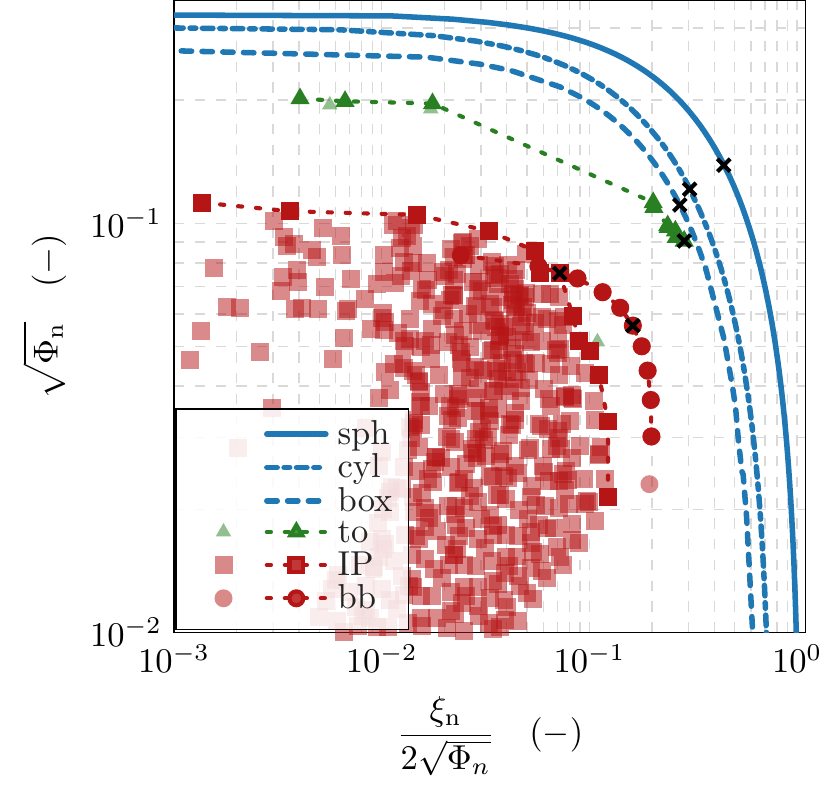}
    \caption{Pareto-optimal points in curvature and a bias magnetic field for different current supporting regions (spherical shell~=~sph, cylindrical shell~=~cyl, box) compared to selected realizations (topology optimization~\cite{Capeketal_ShapeSynthesisBasedOnTopologySensitivity} designs~=~to, Ioffe-Pritchard trap~=~IP, baseball trap~=~bb). The Pareto-optimal points of realizations are highlighted by bold symbols and interconnected by dotted lines. Black crosses denote values for which the current densities are shown in Figs.~\ref{fig:boundCurrents}~--~\ref{fig:IoffePritchardParams},~\ref{fig:topoOpt}.}
    \label{fig:trappingPareto}
\end{figure}

The solution to the optimization problem~\eqref{eq:optim} is approached by its transformation into a convex \ac{QCQP} that is solved by standard methods~\cite{NocedalWright_NumericalOptimization, BoydVandenberghe_ConvexOptimization}. The details are shown in Appendix~\ref{ap:opt}. The Pareto-optimal set of points given by solutions to the optimization problem for the given current supporting regions (a spherical shell, a cylindrical shell, and a box) are shown in Fig.~\ref{fig:trappingPareto}, after applying normalization
\begin{align}
    \Phi_\T{n} &= \dfrac{\rho^2 R_\T{c}}{\MUE^2 P_\mathrm{L}} \Phi, \label{eq:normalizationPot} \\
    \xi_\T{n} &= \dfrac{\rho^4 R_\T{c}}{\MUE^2 P_\mathrm{L}} \xi_\T{min}. \label{eq:normalizationConvex}
\end{align}
The squared magnitude of magnetic field $\Phi$ and the smallest eigenvalue $\xi_\T{min}$ are normalized by radius $\rho$ of the smallest inscribed sphere centered at the trapping center~\footnote{Inscribed sphere radius is determined by shortest distance between the trapping center and any part of conductor forming the trap. The radius is related to the physical dimensions and the trapping volume.}, conductor resistivity~$R_\T{c}$, free space permeability~$\MUE$, and lost power~$P_\T{L}$. Normalization allows geometries to be compared regardless of the physical dimensions or the materials used as conductors. This normalization was also employed in numerical solution to~\eqref{eq:optim} making it scale-invariant. Consulting definitions~\eqref{eq:potTerm0}--\eqref{eq:potTerm2}, the quantities depicted on the axis of Fig.~\ref{fig:trappingPareto} are (apart from linear multiplication by the magnitude of the magnetic dipole moment~$m$) the normalized potential in the trapping center~$\phi$ and potential convexity~$\lambda_\T{min}$.

Comparing current supporting regions with the above-mentioned normalization, the spherical shell performs best in both observed parameters, followed by the cylindrical shell and box. This ordering is nevertheless normalization dependent, in this case, the dependency is due to radius of the largest inscribed sphere~$\rho$, which is native normalization for a spherical current supporting region.

\begin{figure}
    \centering
    \includegraphics[width = 8.6cm]{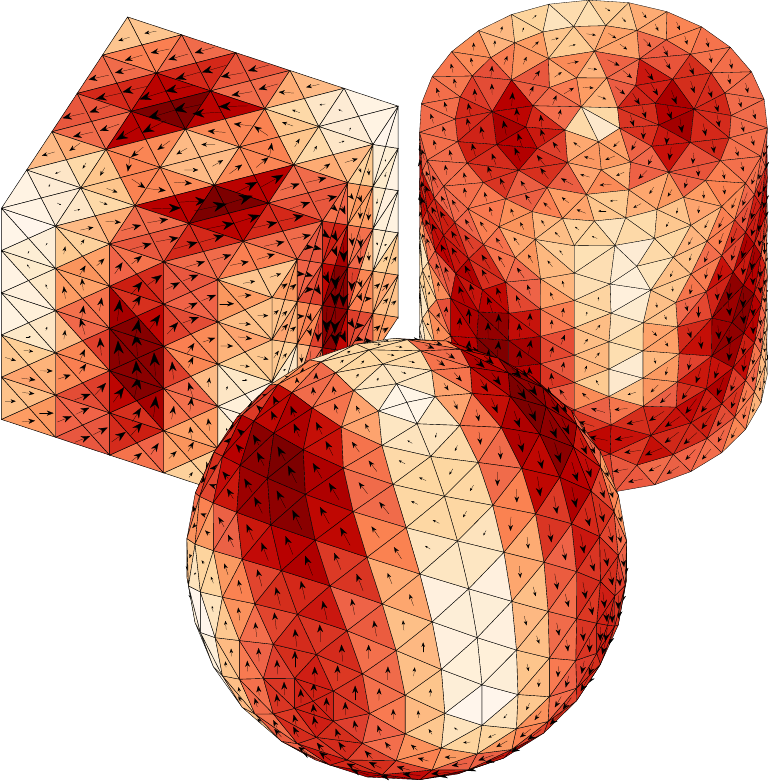}
    \caption{Optimal current densities on the current supporting regions (a spherical shell, a cylindrical shell, and a box) performing according to black cross marks in~Fig.~\ref{fig:trappingPareto}.}
    \label{fig:boundCurrents}
\end{figure}

\begin{figure}
    \centering
    \includegraphics[width = 5cm]{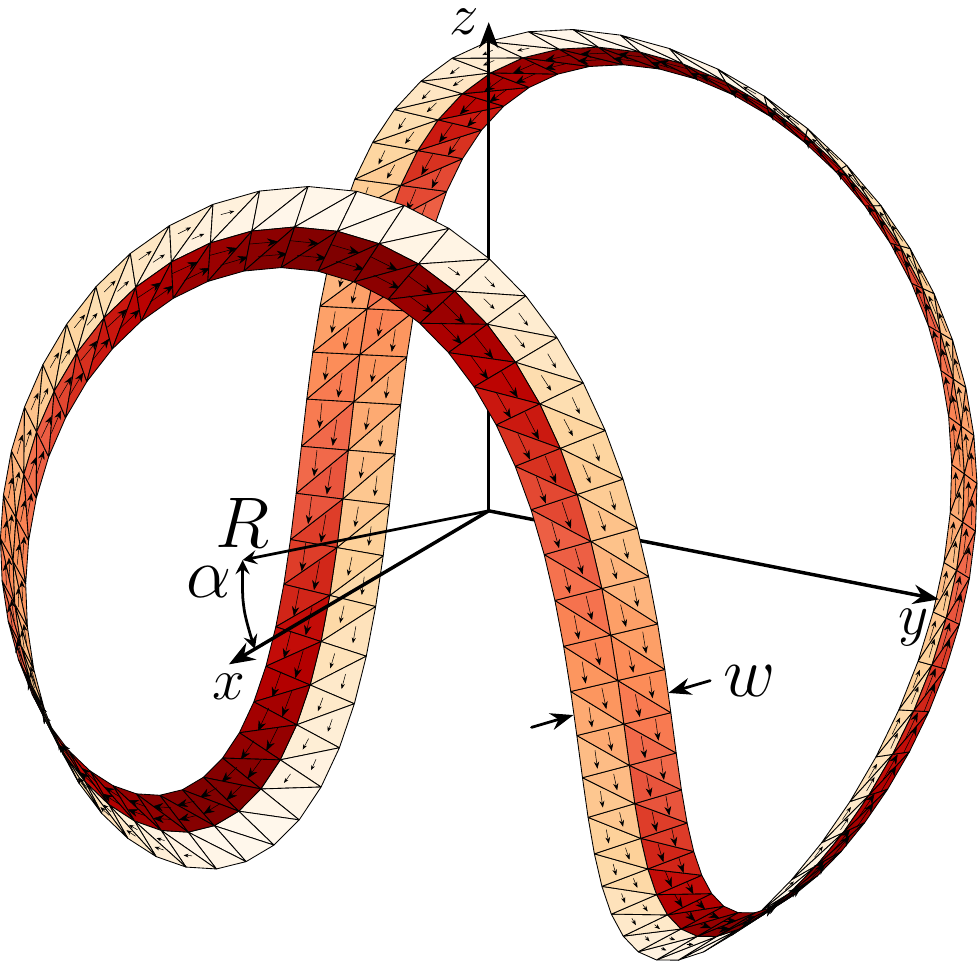}
    \caption{Current density on the baseball trap with angle $\alpha = 22^\circ$. The baseball trap is composed of four contiguous planar arcs. Angle~$\alpha$ measures the tilt of the arc's center with respect to its axis. The four arcs are therefore defined by angles $+\alpha, - \alpha, + \alpha$, and $-\alpha$ with respect to $x, y, - x$, and $-y$ respectively~\cite{Bergeman_MagStatTrapFielsNeutralAtoms}.}
    \label{fig:bsbl22}
\end{figure}

Selected optimal current densities corresponding to black cross marks in~Fig.~\ref{fig:trappingPareto} are shown in Fig.~\ref{fig:boundCurrents}.
The optimal current density changes along the Pareto-optimal set. The optimal points with strong bias magnetic field and low curvature are realized by current densities whose shape resembles a solenoid. This gives an explanation for the lower performance in the bias magnetic field of the baseball trap discussed later because its geometry does not support this kind of current density. The optimal points with the highest curvature are realized by current density resembling a hexapole trap. The central region of the Pareto frontier is occupied by current densities resembling the shape of a baseball's seam, one of the classical trap designs described in Fig.~\ref{fig:bsbl22}. Notice also that the optimal current densities are invariant with respect to the~$\sigma_z C_{4z}$ operation (using Sch\"{o}enflies notation~\cite{McWeeny_GroupTheory}). Later on, this symmetry property is employed in topology optimization.

\begin{figure}
    \centering
    \includegraphics[width = 5cm]{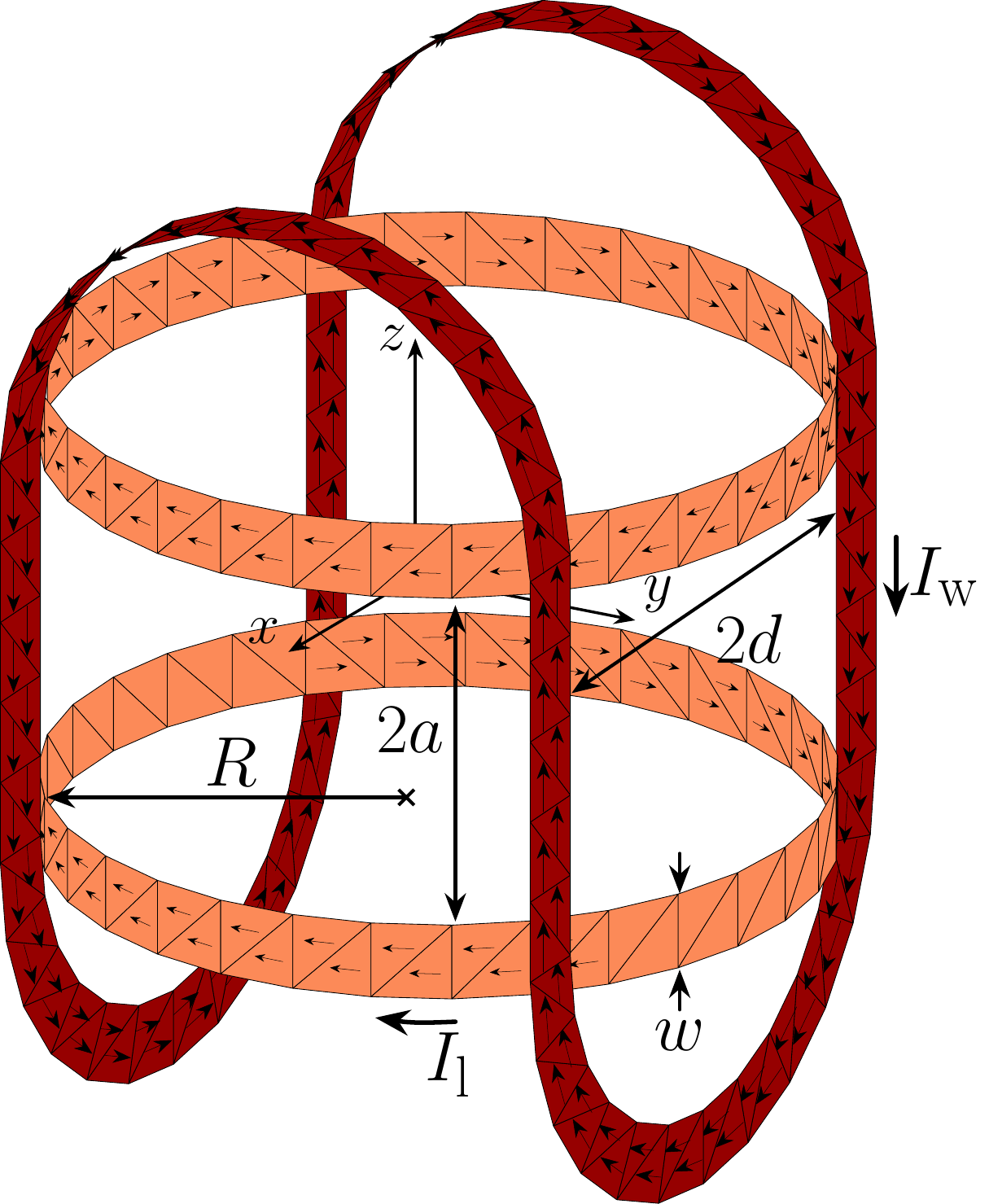}
    \caption{Ioffe-Pritchard trap configuration with performance according to black cross mark  in Fig.~\ref{fig:trappingPareto}. Important variables related to the normalized performance of this trap are ratio of distance between loops and their radius $2a/R = 1.09$, ratio of distance between wires and radius of loops $2d/R = 1.53$, and ratio of currents in loops and wires $I_\T{W}/I_\T{l} = 1.84$.}
    \label{fig:IoffePritchardParams}
\end{figure}

Optimal current densities are typically impossible to realize with realistic excitation. Their most important quality is the performance limitation, which cannot be overcome by any realization. In that respect, Fig.~\ref{fig:trappingPareto} also shows the performance of two widely used trap designs, the baseball trap~\cite{BaseballMagneticField1966,Hiskes_WhoMadeBaseball,Bergeman_MagStatTrapFielsNeutralAtoms} and the Ioffe-Pritchard trap~\cite{Bergeman_MagStatTrapFielsNeutralAtoms,PerezRios_HowMagTrapWork}. The performance of the topology optimized designs is also included. The baseball trap, see~Fig.~\ref{fig:bsbl22}, is parametrized by its angle, in~\cite{Bergeman_MagStatTrapFielsNeutralAtoms} denoted as~$\alpha$.
The Ioffe-Pritchard trap is parametrized by three parameters: the relative radius of the loops creating the bias magnetic field in the trapping center and their distance from the trapping center, the relative distance of the vertical conductors from the trapping center, and the ratio of currents in loops and vertical conductors. The specific design of the Ioffe-Pritchard trap is shown in~Fig.~\ref{fig:IoffePritchardParams}.
Both designs are made of conductive strips of width~$w$. Particularly, strip width~$w = 0.2 R$ is used for all baseball and Ioffe-Pritchard traps with~$R$ being the radius, see~Figs.~\ref{fig:bsbl22} and~\ref{fig:IoffePritchardParams}.

By changing the parameters of the designs, their Pareto-optimal sets were found. The angle defining the baseball trap is Pareto-optimal in the approximate range of $17^\circ$ to $26^\circ$. The Pareto-optimal performances of the Ioffe-Pritchard trap are highlighted in Fig.~\ref{fig:trappingPareto} by bold symbols.

\begin{figure}
    \centering
    \includegraphics{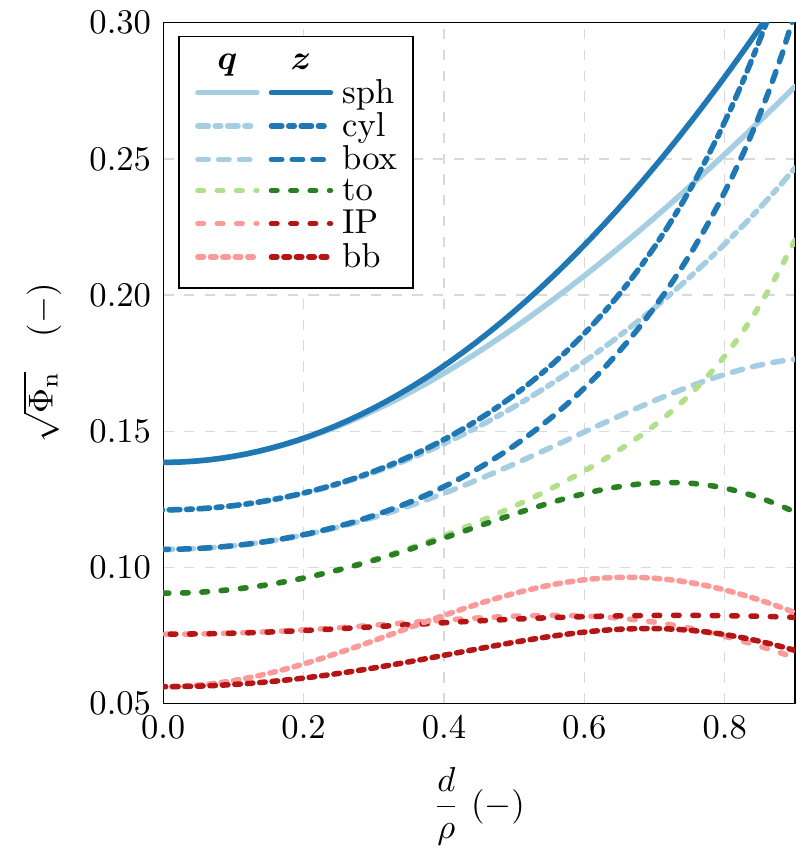}
    \caption{Potential profiles generated by optimal current densities corresponding to maximal curvature. Potential wells are depicted in chosen directions with~$d$ being the distance from the trapping center. Potentials along vector $\V{q}$ are denoted by light lines, while the potentials along $z$ axis are denoted by dark lines.}
    \label{fig:potWellBound}
\end{figure}

Potential wells generated by the realized and optimal current densities corresponding to the black cross marks in Fig.~\ref{fig:trappingPareto} are shown in Fig.~\ref{fig:potWellBound}. Particular vectors~$\V{q}$ along which the potential is depicted are chosen as follows. The potential is identical in plane $xy$ for the spherical shell and the cylindrical shell current supporting regions and arbitrary vector~$\V{q}$ in plane $xy$ can be chosen for them. The potential corresponding to the box support is shown along vector $\V{v}^\trans = [1,1,0]$, which is the cut of the overall lowest increase of potential from the center in the $xy$ plane. The depicted lines, nevertheless, show that the potential realized by the optimal current densities is almost isotropic if $d/\rho < 0.4$, where the potential is identical in all directions, and where the quadratic approximation is sufficient. Potential wells realized by current densities in the baseball and Ioffe-Pritchard traps, as well as the results of topology optimization, are not generally isotropic and the potential is depicted along the $x$ and $z$ axes, which do not intersect with any conducting part of the trap. Although, in general, higher curvature at the trapping center does not ensure an overall better trapping depth, the curves in Fig.~\ref{fig:potWellBound} show that it is the case for the optimal current density. In the sense of fundamental bounds, the local quadratic approximation is, therefore, legitimate.
Despite the fact that the realized trap designs are not forced to form locally isotropic potential wells, Fig.~\ref{fig:trappingPareto} shows a large gap between the realized performance and the fundamental bounds. This gap calls for better trap designs.

The normalized performance of the realized traps depends on the width of the strip, which is mainly related to ohmic losses. By widening the conducting strips, performance improvement can be achieved. This, nevertheless, has diminishing returns since current density is concentrated at the edges of the current carrying region. The simple idea of dividing the strip into several disconnected current paths also does not lead to any significant improvement. These observations show the necessity of topology changes which are, in this paper, performed by the topology optimization scheme developed in~\cite{Capeketal_ShapeSynthesisBasedOnTopologySensitivity}. Topology optimized designs can also verify the feasibility of performance limitations.

\begin{figure}
    \centering
    \includegraphics[width = 5cm]{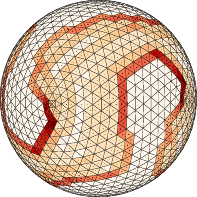}
    \caption{Current density excited in design given by topology optimization and corresponding to the black cross mark in Fig.~\ref{fig:trappingPareto}. As compared to the baseball trap in Fig.~\ref{fig:bsbl22}, the topology optimized design achieves almost two times higher normalized potential curvature.}
    \label{fig:topoOpt}
\end{figure}

In topology optimization, the objective function is a convex combination of field strength and potential convexity. The vanishing force in the trapping center is ensured by assuming the geometry and feeding invariant with respect to the~$\sigma_z C_{4z}$ operation~\cite{McWeeny_GroupTheory}, see Appendix~\ref{ap:opt}. Optimization is applied to a spherical shell fed by four identical sources placed at the maxima of the optimal current density shown in Fig.~\ref{fig:boundCurrents}. The performance of Pareto-optimal designs obtained by topology optimization is shown in Fig.~\ref{fig:trappingPareto}. Furthermore, Fig.~\ref{fig:topoOpt} shows the design corresponding to black cross mark in Fig.~\ref{fig:trappingPareto}. The shape of the potential well generated by the current through this trap is included in Fig.~\ref{fig:potWellBound}.
Qualitatively, topology optimization tries to reassemble the optimal current density given by the fundamental bound on the spherical shell in Fig.~\ref{fig:boundCurrents}. The relative performance, as compared to fundamental bound, depends on the number of excitation points and their positioning in the supporting region.

The performance of topologically optimized designs is significantly higher than that of the baseball and Ioffe-Pritchard traps. The improvement indicates that the hunt for the optimal trap design is open for further investigation. In addition, the results of topology optimization show that the requirement on fundamental bounds to form a locally isotropic potential well is natural, since the topologically optimized designs also form a locally isotropic potential well, despite not being forced to do so.

Despite the positive outcomes shown above, a further investigation of the topic is nevertheless needed. The topologically optimized design, as well as the design of Ioffe-Pritchard and baseball traps shown in this paper, all omit many practical design aspects, such as realistic feeding and mechanical support of trap conductors. Furthermore, the topology optimization used in this paper~\cite{Capeketal_ShapeSynthesisBasedOnTopologySensitivity} relies on addition or removal of basis functions~$\basisFcn_i$ defined in Appendix~\ref{ap:MoM}. The optimized design must, therefore, be interpreted only as a set of chosen basis functions, which creates the optimized current path. Such a structure is complicated to fabricate because it is the interconnection of neighbor elements and not the presence of mesh elements which is optimized.

\section{Addition of Gravity}

The preceding text did not consider the effects of gravity. In cases of light particles and a strong prepolarizing field, the previous theory can still be used, and only after optimization, the magnitude of the current is set so as to compensate for gravity at a slightly off-center location. This process in general breaks the invariance of the results to absolute dimensions and physically larger traps suffer less from the effects of gravity. In cases of physically small traps and heavy particles, the effect of gravity must be taken into account rigorously. This section shows the necessary modifications to the already developed theoretical framework.

A starting point is an addition of a linear term into total potential energy
\begin{equation}
    \phi_\T{t} = - \V{m} \cdot \V{B} - m_\T{g} \V{g} \cdot \V{r}.
\end{equation}
The gravitational field is assumed to be homogeneous and dependent on object mass~$m_\T{g}$ and gravity vector~$\V{g}$. Proceeding further to the local approximation~\eqref{eq:Taylor}, the linear term is preserved
\begin{equation}
    \phi_\T{t} = \phi_\T{0} - \left( \V{F} + m_\T{g} \V{g} \right) \cdot \V{r} + \frac{1}{2} \V{r} \cdot \M{H} \cdot \V{r}.
\end{equation}
 Using auxiliary quantities quadratic in magnetic field~\eqref{eq:potB}~--\eqref{eq:hessB}, the terms in the Taylor polynomial read the same as~\eqref{eq:potTerm0}--\eqref{eq:potTerm2} except the total force, which incorporates the gravitational force. The resulting multi-objective optimization equivalent to~\eqref{eq:optim} reads
 \begin{equation}\label{eq:optimG}
\begin{split}
    \max \limits_{\V{J}} \ & \lambda_\T{min} \\
    \subto \ & P_\T{L} = P_\T{L}^\T{max}, \\
    & \Phi = \Phi_\T{s}, \ \Phi_\T{s} \in [\Phi_\T{min}, \Phi_\T{max}], \\
    & F_i = -m_\T{g} g_i, \ i \in \{x,y,z\}.
\end{split}
\end{equation}
As in the gravity-free case, the solution to the optimization problem~\eqref{eq:optimG} is provided by its transformation into a convex \ac{QCQP}. Details can be found in Appendix~\ref{ap:gravity}.

Consideration of gravity imposes an increase in computational complexity and loss of generality, which are also highlighted in Appendix~\ref{ap:gravity}. If gravity is taken into account, the maximal allowed lost power, mass, magnetic dipole, and gravity have to be explicitly specified. The results thus become particle-dependent and are not independent of absolute dimensions anymore. The detailed study of gravity effects is left as a topic of future, more design-focused, research.

\section{Conclusion}
The performance limits on the magnetic traps were studied using local approximation and numerical methods, mainly the Galerkin method and convex optimization. The bias field and convexity were the analyzed objectives. Fundamental bounds of normalized metrics, the Pareto-optimal set, were found for different current supporting regions: a spherical shell, a cylindrical region, and a box. Performances of the baseball trap and the Ioffe-Pritchard trap were compared to the limits given by the regions, which supports these designs.

Optimal current density realizing the fundamental bound can serve as an inspiration for the improvement of the actual designs. Nevertheless, shape optimization was also used to narrow the gap in the performance of fundamental bounds and known designs. Design topology is optimized in order to increase its performance with the absolute benchmark being the fundamental bound for the same current supporting the region.

The determination of fundamental bounds is an important step when the performance of a design is assessed and future research in this area should focus on the extension of the presented formulation for magnetic traps with multiple trapping centers~\cite{HuYin_ControllableDoubleWellMagTraps}, which can easily be treated solely by modifying the objective function and by adding constraints. Besides, a detailed study should be performed on the effects of gravity on specific particles and their influence significance on performance limitations. In addition, an effort should also be made in extending the formulation into the fields of trapping charged particles, \ie, plasma confinement. 

\begin{acknowledgments}
We would like to acknowledge the financial support of this work by the Czech Science Foundation under project~\mbox{No.~21-19025M}.
\end{acknowledgments}

\appendix

\section{Matrix Description}\label{ap:MoM}
The description of field in this paper is based on current density expansion into a set of basis functions~$\basisFcn_i$. Current density flowing in conductors is represented by these basis functions with expansion coefficients~$I_i$,
\begin{equation}\label{eq:expansion}
\V{J} (\V{r}) \approx \sum \limits_i I_i \basisFcn_i (\V{r})
\end{equation}
collected into column vector $\Ivec = [I_i]$. In the case of optimization, current expansion vector~$\Ivec$ is taken as the optimized variable~\cite{GustafssonTayliEhrenborgEtAl_AntennaCurrentOptimizationUsingMatlabAndCVX,GustafssonCismasuJonsson_PhysicalBoundsAndOptimalCurrentsOnAntennas_TAP,JelinekCapek_OptimalCurrentsOnArbitrarilyShapedSurfaces}. In the case of trap design, the current expansion vector results from a solution to a system of linear equations
\begin{equation}
    \Zmat \Ivec = \Vvec,
\end{equation}
where $\Zmat$ is the system (impedance) matrix and $\Vvec$ collects the projections of electric excitation onto the used basis functions. In this paper, system matrix~$\M{Z}$ comes from the application of the Galerkin method~\cite{Kantorovich1982} or~\acf{MoM}~\cite{Harrington_FieldComputationByMoM,Gibson_MoMinElectromagnetics}, to the electric field integral equation. Loop-star decomposition~\cite{Vecch_iLoopStar} has been employed to allow the operation under quasistatic conditions. In all studied cases, the current density was composed of loop-like currents which assures a vanishing divergence of current density. The decomposition is performed numerically with high precision, considering the electric energy to be vanishingly small as compared to magnetic energy.

Expansion~\eqref{eq:expansion} also allows a magnetic field with its spatial derivatives and lost power to be expressed as linear or quadratic forms of column vector~$\Ivec$:
\begin{itemize}
    \item The magnetic field and its derivatives at the trapping center are represented by matrices $\M{B}$, $\M{B}_{,i}$, and $\M{B}_{,ij}$ as
    \begin{align}
        \V{B} &\approx \M{B} \Ivec, \label{eq:Bmat} \\
        \rho \frac{\partial \V{B}}{\partial i} &\approx \M{B}_{,i} \Ivec, \label{eq:BmatDer1} \\
        \rho^2 \frac{\partial^2 \V{B}}{\partial j \partial i} &\approx \M{B}_{,ij} \Ivec, \label{eq:BmatDer2}
    \end{align}
    respectively, where $i$ and $j$ represent spatial variables $x,y,z$ and comma in the lower index stands for derivatives with respect to variables following it. Spatial derivatives of magnetic field are normalized by the shortest distance~$\rho$ between the trapping center and the current carrying region. Such normalization enforces equal units to all matrices~$\M{B}$, $\M{B}_{,i}$, and $\M{B}_{,ij}$ which is advantageous for mathematical operations in Appendix~\ref{ap:opt}.
    \item Lost power as
    \begin{equation}
    P_\T{L} \approx \frac{1}{2} \Ivec^\herm \Rmat_\rho \Ivec,
    \end{equation}
    where $\Rmat_\rho$ is the lost power matrix~\cite{Harrington_FieldComputationByMoM,JelinekCapek_OptimalCurrentsOnArbitrarilyShapedSurfaces}.
\end{itemize}
According to~\eqref{eq:potB}--\eqref{eq:hessB}, the matrices of the magnetic field and its spatial derivatives~\eqref{eq:Bmat}--\eqref{eq:BmatDer2} are used to express the squared magnitude of the magnetic field, its gradient's elements, and the Hessian matrix's elements as quadratic forms:
\begin{align}
    \Phi &\approx \Ivec^\herm \M{\Phi} \Ivec, \\
    \rho \Gamma_i &\approx \Ivec^\herm \M{\Gamma}_i \Ivec, \\
    \rho^2 \Xi_{ij} &\approx \Ivec^\herm \M{\Xi}_{ij} \Ivec,
\end{align}
where
\begin{align}
    \M{\Phi} &= \M{B}^\herm \M{B}, \\
    \M{\Gamma}_i & = \M{B}_{,i}^\herm \M{B} + \M{B}^\herm \M{B}_{,i},\\
    \M{\Xi}_{ij} &= \M{B}_{,ij}^\herm \M{B} + \M{B}^\herm \M{B}_{,ij} + \M{B}_{,i}^\herm \M{B}_{,j} + \M{B}_{,j}^\herm \M{B}_{,i},
\end{align}
all share the same units.

\section{Convex Optimization}\label{ap:opt}
Evaluation of fundamental bounds on magnetic confinement is based on optimization where the current expansion coefficients~\eqref{eq:expansion} are the degrees of freedom~\eqref{eq:optim}. The optimization is formulated as \acf{QCQP}~\cite{NocedalWright_NumericalOptimization,BoydVandenberghe_ConvexOptimization}.

In order to rewrite \eqref{eq:optim} as \ac{QCQP}, an assumption on the Hessian matrix~\eqref{eq:hessB} has to be made. The minimal eigenvalue~\eqref{eq:minEig} in \eqref{eq:optim} cannot, in general, be represented via a quadratic function. To overcome this difficulty, the off-diagonal terms in the Hessian matrix are enforced to be zero and the diagonal terms are enforced so all are equal by the constraints of optimization. The restriction results in a locally isotropic potential well in the magnetic trap and, due to the symmetry, it can be applied to a spherical current supporting region without any impact. The same assumption is also done for other current supporting regions: cylindrical shell and box. In these later cases, such a bound might then be too pessimistic as better performance can possibly be achieved by using a non-symmetrical potential well. The results presented in Sec.~\ref{sec:res}, nevertheless, show that common trap designs still perform worse than these ``restricted'' fundamental bounds.

The \ac{QCQP} equivalent to the optimization problem~\eqref{eq:optim} with the aforementioned assumption reads
\begin{equation}
\begin{split}
    \max \limits_\Ivec \quad & \Ivec^\herm \M{\Xi}_{xx} \Ivec, \\
    \T{s.t.} \quad & \Ivec^\herm \Rmat_\rho \Ivec - 2P_\T{L}^\T{max} = 0, \\
    & \Ivec^\herm \left[ \M{\Phi} - \dfrac{\Phi_\T{s}}{2P_\T{L}^\T{max}} \Rmat_\rho \right] \Ivec = 0, \ \Phi_\T{s} \in [\Phi_\T{min}, \Phi_\T{max}]\\
    & \Ivec^\herm \M{\Gamma}_i \Ivec = 0, \ i \in \{x,y,z\}\\
    & \Ivec^\herm (\M{\Xi}_{ii} - \M{\Xi}_{jj}) \Ivec = 0, \ \forall i \neq j \\
    & \Ivec^\herm \M{\Xi}_{ij} \Ivec = 0, \ \forall i \neq j.
\end{split}
\label{eq:optim:QCQP}
\end{equation}
The problem is numerically solved via a dual formulation~\cite{BoydVandenberghe_ConvexOptimization, NocedalWright_NumericalOptimization} by using MATLAB code~\cite{Liska_etal_FundamentalBoundsEvaluation} and exhibits no duality gap for the studied scenarios. Numerical implementation should also include normalizations~\eqref{eq:normalizationPot},~\eqref{eq:normalizationConvex} which make it scale-invariant. In order to demonstrate the solution and implementation steps, a simplified MATLAB script is accessible as ``exMagTrap.m'' in the examples folder of Fundamental Bounds addon at \url{http://antennatoolbox.com/fundamentalBounds}.

Although the problem~\eqref{eq:optim}, and therefore~\eqref{eq:optim:QCQP}, can be solved directly, it allows for a large reduction in the degrees of freedom which is computationally advantageous. From~\eqref{eq:optim}, with~\eqref{eq:potB} -- \eqref{eq:hessB} in mind, it is observed that the current density generating a simultaneously vanishing magnetic field, as well as its first and second derivatives, cannot contribute to the solution. The appropriate low dimensional solution subspace is, therefore, given by eigenvectors corresponding to nonzero eigenvalues of a \ac{GEP}
\begin{equation}\label{eq:gepOmega}
\left( \M{\Omega}^\herm \M{\Omega} \right) \Ivec = \beta \Rmat_\rho \Ivec,
\end{equation}
where
\begin{equation}
    \M{\Omega} = \begin{bmatrix}
    \M{B}^\herm &
    \M{B}_{,x}^\herm &
    \M{B}_{,y}^\herm &
    \M{B}_{,z}^\herm &
    \M{B}_{,xx}^\herm &
    \M{B}_{,xy}^\herm &
    \hdots &
    \M{B}_{,zz}^\herm
    \end{bmatrix}^\herm.
\end{equation}
Due to the orthogonality of the eigenvectors with respect to lost power, those eigenvectors with zero eigenvalues can only decrease the performance of the magnetic trap.

If a magnetic field is considered only in the direction along the $z$-axis, as is the case of Sec.~\ref{sec:res}, the force (the third constraint in~\eqref{eq:optim}) can be zeroed \textit{a priori} before solving the optimization problem. This is done by using the null space of a matrix, the rows of which are specific components of a magnetic field operator and of its first derivatives, as a solution subspace. Namely, components $\M{B}_x,\M{B}_y$ and $\M{B}_{z,i}, \ \forall i \in \{x,y,z\}$ are used in Sec.~\ref{sec:res}. Consulting the law of Biot and Savart~\eqref{eq:BiotSavart}, members of this subspace are invariant with respect to the~$\sigma_z C_{4z}$ operation~\cite{McWeeny_GroupTheory}, whenever the underlying geometry allows it. This is advantageously used in topology optimization, where the constraint on zero force in the trapping center is enforced by aforementioned symmetry of structure and feeding, greatly reducing the optimization complexity. \\

\section{Computational Aspects of Gravity Addition}\label{ap:gravity}
Following~\eqref{eq:optimG} to find the fundamental bound on the magnetic trap and Appendix~\ref{ap:opt}, changes need to be done in comparison to~\eqref{eq:optim:QCQP} in order to incorporate non-vanishing magnetic force at the trapping center. The modified optimization problem (compare to \eqref{eq:optim:QCQP}) reads
 \begin{equation} \label{eq:optimG:QCQP} \begin{split}
     \max \limits_\Ivec \quad & \Ivec^\herm \M{\Xi}_{xx} \Ivec, \\
    \T{s.t.} \quad & \Ivec^\herm \Rmat_\rho \Ivec - 2P_\T{L}^\T{max} = 0, \\
    & \Ivec^\herm \left[ \M{\Phi} - \dfrac{\Phi_\T{s}}{2P_\T{L}^\T{max}} \Rmat_\rho \right] \Ivec = 0, \ \Phi_\T{s} \in [\Phi_\T{min}, \Phi_\T{max}]\\
    & \Ivec^\herm \M{\Gamma}_i \Ivec - \dfrac{2 m_\T{g} g_i \sqrt{\Phi_\T{s}}}{m} = 0,\\
    & \Ivec^\herm \left(\M{\Xi}_{ii} - \M{\Xi}_{jj} \right) \Ivec - \dfrac{2 m_\T{g}^2}{m^2} \left(g_i^2 - g_j^2 \right) = 0, \\
    & \Ivec^\herm \M{\Xi}_{ij} \Ivec - \dfrac{2 m_\T{g}^2 g_i g_j}{m^2}= 0, 
 \end{split} \end{equation}
where $i,j$ span all spatial coordinates $\{x,y,z\}$. The solution is implicitly dependent on parameters $P_\T{L}^\T{max}, \ m_\T{g}, \ g_i, \ m$, which determine all possible Pareto-optimal solutions. The optimization can be solved by using the same tool as in the gravity-free case.

Advantageously, \eqref{eq:gepOmega} can still be applied in order to decrease the complexity of the computation. Moreover, if a magnetic field and gravity are considered only in the direction along the $z$-axis, the force (the third constraint in~\eqref{eq:optimG:QCQP}) can be treated \textit{a priori} as in Appendix~\ref{ap:opt}. The only difference is that matrix~$\M{B}_{z,z}$ is excluded from the matrix whose null space is used to zero the forces in the direction along $x$- and $y$-axis. The component $\M{B}_{z,z}$ is used to oppose the gravity.

Contrary to the gravity-free case, the members of the aforementioned null space are in general not invariant with respect to the~$\sigma_z C_{4z}$ operation and this cannot be subsequently used to simplify topology optimization. In addition, the constraint equalizing the gravitational and magnetic forces has to be treated, which results in a considerable increase in the computational complexity of topology optimization.

\bibliographystyle{ieeetr}
\bibliography{references.bib}

\begin{thebibliography}{10}

\bibitem{Wing_NeutralPartTrapQuasistaElMagFields}
W.~H. Wing, ``On neutral particle trapping in quasistatic electromagnetic
  fields,'' {\em Progress in Quantum Electronics}, vol.~8, pp.~181--199, jan
  1984.

\bibitem{Gov_MagTrapNeutralParticles}
S.~Gov, S.~Shtrikman, and H.~Thomas, ``Magnetic trapping of neutral particles:
  Classical and quantum-mechanical study of a ioffe{\textendash}pritchard type
  trap,'' {\em Journal of Applied Physics}, vol.~87, pp.~3989--3998, apr 2000.

\bibitem{PerezRios_HowMagTrapWork}
J.~P{\'{e}}rez-R{\'{\i}}os and A.~S. Sanz, ``How does a magnetic trap work?,''
  {\em American Journal of Physics}, vol.~81, pp.~836--843, nov 2013.

\bibitem{Chu_ManipulationNeutralPart-Lecture}
S.~Chu, ``Nobel lecture: The manipulation of neutral particles,'' {\em Reviews
  of Modern Physics}, vol.~70, pp.~685--706, jul 1998.

\bibitem{CohenTannoudji_ManipulAtomsWithFotons-Lecture}
C.~N. Cohen-Tannoudji, ``Nobel lecture: Manipulating atoms with photons,'' {\em
  Reviews of Modern Physics}, vol.~70, pp.~707--719, jul 1998.

\bibitem{Phillips_LaserCoolingTrapNetralAtoms_Lecture}
W.~D. Phillips, ``Nobel lecture: Laser cooling and trapping of neutral atoms,''
  {\em Reviews of Modern Physics}, vol.~70, pp.~721--741, jul 1998.

\bibitem{Phillips_LaserCoolElMagTrapNeutralAtoms}
W.~D. Phillips, J.~V. Prodan, and H.~J. Metcalf, ``Laser cooling and
  electromagnetic trapping of neutral atoms,'' {\em Journal of the Optical
  Society of America B}, vol.~2, p.~1751, nov 1985.

\bibitem{Anderson_BoseEinsteinCondDiluteAtomicVapor}
M.~H. Anderson, J.~R. Ensher, M.~R. Matthews, C.~E. Wieman, and E.~A. Cornell,
  ``Observation of bose-einstein condensation in a dilute atomic vapor,'' {\em
  Science}, vol.~269, pp.~198--201, jul 1995.

\bibitem{Davis_BoseEinsteinCondGasNaAtoms}
K.~B. Davis, M.~O. Mewes, M.~R. Andrews, N.~J. van Druten, D.~S. Durfee, D.~M.
  Kurn, and W.~Ketterle, ``Bose-einstein condensation in a gas of sodium
  atoms,'' {\em Physical Review Letters}, vol.~75, pp.~3969--3973, nov 1995.

\bibitem{Bradley_BoseEinsteinCondAtomicGasAttraciveInteract}
C.~C. Bradley, C.~A. Sackett, J.~J. Tollett, and R.~G. Hulet, ``Evidence of
  bose-einstein condensation in an atomic gas with attractive interactions,''
  {\em Physical Review Letters}, vol.~75, pp.~1687--1690, aug 1995.

\bibitem{Bergeman_MagStatTrapFielsNeutralAtoms}
T.~Bergeman, G.~Erez, and H.~J. Metcalf, ``Magnetostatic trapping fields for
  neutral atoms,'' {\em Physical Reviaw A}, vol.~35, pp.~1535--1546, feb 1987.

\bibitem{Pritchard_CoolNeutralAtomsMagTrapPrecSpectroscopy}
D.~E. Pritchard, ``Cooling neutral atoms in a magnetic trap for precision
  spectroscopy,'' {\em Physical Review Letters}, vol.~51, pp.~1336--1339, oct
  1983.

\bibitem{Gott_ResultsConfMagTraps}
Y.~V. Gott, M.~S. Ioffe, and V.~G. Telkovskii, ``Some new results on
  confinement in magnetic traps,'' {\em Nuclear Fusion (Austria)}, vol.~Vol:
  Suppl. 2, Pt. 3, 1 1962.

\bibitem{BaseballMagneticField1966}
R.~F. Post and C.~C. Brown, ``Baseball magnetic field,'' {\em Physics Today},
  vol.~19, pp.~70--71, nov 1966.

\bibitem{Hiskes_WhoMadeBaseball}
J.~R. Hiskes, ``Who made the baseball?,'' {\em Physics Today}, vol.~20,
  pp.~9--10, jan 1967.

\bibitem{Yang_DevelopmentHighFieldSupercondIoffeMagTraps}
L.~Yang, C.~R. Brome, J.~S. Butterworth, {\em et~al.}, ``Invited article:
  Development of high-field superconducting ioffe magnetic traps,'' {\em Review
  of Scientific Instruments}, vol.~79, p.~031301, mar 2008.

\bibitem{Ahokas_LargeOctupoleMagTrapResearchAtomicHydrogen}
J.~Ahokas, A.~Semakin, J.~Järvinen, O.~Hanski, A.~Laptiyenko, V.~Dvornichenko,
  K.~Salonen, Z.~Burkley, P.~Crivelli, A.~Golovizin, V.~Nesvizhevsky, F.~Nez,
  P.~Yzombard, E.~Widmann, and S.~Vasiliev, ``A large octupole magnetic trap
  for research with atomic hydrogen,'' {\em arXiv}, Aug. 2021.

\bibitem{Harris_MagTrapsNearlyUntrappableParticlesDevelopmentHighFieldSupercondIoffeTraps}
J.~G.~E. Harris, ``Perspective: Magnetic traps for nearly untrappable
  particles: {\textquotedblleft}development of high field superconducting ioffe
  traps{\textquotedblright} [rev. sci. instrum. 79, 031301 (2008)],'' {\em
  Review of Scientific Instruments}, vol.~79, p.~030901, mar 2008.

\bibitem{GustafssonTayliEhrenborgEtAl_AntennaCurrentOptimizationUsingMatlabAndCVX}
M.~Gustafsson, D.~Tayli, C.~Ehrenborg, M.~Cismasu, and S.~Norbedo, ``Antenna
  current optimization using {MATLAB} and {CVX},'' {\em {FERMAT}}, vol.~15,
  pp.~1--29, May--June 2016.

\bibitem{GustafssonCismasuJonsson_PhysicalBoundsAndOptimalCurrentsOnAntennas_TAP}
M.~Gustafsson, M.~Cismasu, and B.~L.~G. Jonsson, ``Physical bounds and optimal
  currents on antennas,'' {\em IEEE Trans. Antennas Propag.}, vol.~60,
  pp.~2672--2681, June 2012.

\bibitem{JelinekCapek_OptimalCurrentsOnArbitrarilyShapedSurfaces}
L.~Jelinek and M.~Capek, ``Optimal currents on arbitrarily shaped surfaces,''
  {\em IEEE Trans. Antennas Propag.}, vol.~65, pp.~329--341, Jan. 2017.

\bibitem{2020_Gustafsson_NJP}
M.~Gustafsson, K.~Schab, L.~Jelinek, and M.~Capek, ``Upper bounds on absorption
  and scattering,'' {\em New Journal of Physics}, vol.~22, p.~073013, sep 2020.

\bibitem{2021_Jelinek_OPEX}
L.~Jelinek, M.~Gustafsson, M.~Capek, and K.~Schab, ``Fundamental bounds on the
  performance of monochromatic passive cloaks,'' {\em Optics Express}, vol.~29,
  no.~15, pp.~24068--24082, 2021.

\bibitem{Venkataram_FundamentalLimitsToRadiativeHeatTransferPRL}
P.~S. Venkataram, S.~Molesky, W.~Jin, and A.~W. Rodriguez, ``Fundamental limits
  to radiative heat transfer: The limited role of nanostructuring in the
  near-field,'' {\em Phys. Rev. Lett.}, vol.~124, p.~013904, Jan 2020.

\bibitem{Molesky_FundamentalLimitsToRadiativeHeatTransferPRB}
S.~Molesky, P.~S. Venkataram, W.~Jin, and A.~W. Rodriguez, ``Fundamental limits
  to radiative heat transfer: Theory,'' {\em Phys. Rev. B}, vol.~101,
  p.~035408, Jan 2020.

\bibitem{2021_Chao_Arxiv}
P.~Chao, B.~Strekha, R.~K. Defo, S.~Molesky, and A.~W. Rodriguez, ``Physical
  limits on electromagnetic response,'' {\em arXiv preprint arXiv: 2109.05667},
  2021.

\bibitem{ChewTongHu_IntegralEquationMethodsForElectromagneticAndElasticWaves}
W.~C. Chew, M.~S. Tong, and B.~Hu, {\em Integral Equation Methods for
  Electromagnetic and Elastic Waves}.
\newblock Morgan \& Claypool, 2009.

\bibitem{Harrington_TimeHarmonicElmagField}
R.~F. Harrington, {\em Time-Harmonic Electromagnetic Fields}.
\newblock Wiley -- IEEE Press, 2~ed., 2001.

\bibitem{Schwinger_ClassicalElectrodynamics}
J.~Schwinger, L.~L. DeRaad, K.~A. Milton, and T.~W.-y., {\em Classical
  Electrodynamics}.
\newblock Westview Press, 1998.

\bibitem{NocedalWright_NumericalOptimization}
J.~Nocedal and S.~Wright, {\em Numerical Optimization}.
\newblock New York, United States: Springer, 2006.

\bibitem{BoydVandenberghe_ConvexOptimization}
S.~Boyd and L.~Vandenberghe, {\em Convex Optimization}.
\newblock Cambridge, Great Britain: Cambridge University Press, 2004.

\bibitem{Gov_1DtoyModelMagTrapping}
S.~Gov, S.~Shtrikman, and H.~Thomas, ``1{D} toy model for magnetic trapping,''
  {\em American Journal of Physics}, vol.~68, pp.~334--343, apr 2000.

\bibitem{Zangwill_Modern_Electrodynamics}
A.~Zangwill, {\em Modern Electrodynamics}.
\newblock Cambridge University Press, 2012.

\bibitem{Weinstein_MicroscopicMagneticTrapsNeutralAtoms}
J.~D. Weinstein and K.~G. Libbrecht, ``Microscopic magnetic traps for neutral
  atoms,'' {\em Physical Review A}, vol.~52, pp.~4004--4009, nov 1995.

\bibitem{1978CohonMultiobjectiveProgrammingAndPlanning}
L.~J. Cohon, {\em Multiobjective Programming and Planning}.
\newblock Academic Press, 1st~ed., 1978.

\bibitem{Kantorovich1982}
L.~V. Kantorovich and G.~P. Akilov, {\em Functional analysis}.
\newblock Oxford New York: Pergamon Press, 1982.

\bibitem{Harrington_FieldComputationByMoM}
R.~F. Harrington, {\em Field Computation by Moment Methods}.
\newblock Piscataway, New Jersey, United States: Wiley -- IEEE Press, 1993.

\bibitem{Gibson_MoMinElectromagnetics}
W.~C. Gibson, {\em The Method of Moments in Electromagnetics}.
\newblock Chapman and Hall/CRC, 2~ed., 2014.

\bibitem{SenoirVolakis_ApproximativeBoundaryConditionsInEM}
T.~B.~A. Senior and J.~L. Volakis, {\em Approximate Boundary Conditions in
  Electromagnetics}.
\newblock IEE, 1995.

\bibitem{Note1}
Especially volumetric current supporting regions suffer from a large number of
  degrees of freedom and become extremely computationally demanding in the case
  of topology optimization.

\bibitem{Capeketal_ShapeSynthesisBasedOnTopologySensitivity}
M.~Capek, L.~Jelinek, and M.~Gustafsson, ``Shape synthesis based on topology
  sensitivity,'' {\em IEEE Trans. Antennas Propag.}, vol.~67, pp.~3889 -- 3901,
  June 2019.

\bibitem{Note2}
Inscribed sphere radius is determined by shortest distance between the trapping
  center and any part of conductor forming the trap. The radius is related to
  the physical dimensions and the trapping volume.

\bibitem{McWeeny_GroupTheory}
R.~McWeeny, {\em Symmetry: An Introduction to Group Theory and Its
  Applications}.
\newblock London: Pergamon Press, 1963.

\bibitem{HuYin_ControllableDoubleWellMagTraps}
J.~Hu and J.~Yin, ``Controllable double-well magnetic traps for neutral
  atoms,'' {\em Journal of the Optical Society of America B}, vol.~19, p.~2844,
  dec 2002.

\bibitem{Vecch_iLoopStar}
G.~Vecchi, ``Loop-star decomposition of basis functions in the discretization
  of the {EFIE},'' {\em {IEEE} Transactions on Antennas and Propagation},
  vol.~47, no.~2, pp.~339--346, 1999.

\bibitem{Liska_etal_FundamentalBoundsEvaluation}
J.~Liska, L.~Jelinek, and M.~Capek, ``Fundamental bounds to time-harmonic
  quadratic metrics in electromagnetism: {O}verview and implementation,'' {\em
  arXiv}, 2021.

\end{thebibliography}
\end{document}